\documentclass[twocolumn,showpacs,preprintnumbers,amsmath,amssymb,aps,pre]{revtex4}

\usepackage{latexsym}
\usepackage{dcolumn}
\usepackage{supertabular,multirow}
\usepackage{epsfig}
\usepackage{bm}
\usepackage{subfigure}
\usepackage{color}
\usepackage{hyperref}
\usepackage{tikz,siunitx}
\usepackage{amssymb,amsfonts,amsmath}
\usepackage{euscript}
\usepackage{graphics}
\usepackage{setspace}
\usepackage{graphicx}
\usepackage{hyperref}
\usepackage{xcolor}
\usepackage{siunitx}
\usepackage[normalem]{ulem}

\definecolor{brown}{rgb}{0.59, 0.29, 0.0}
\definecolor{orange}{RGB}{255,127,0}
\definecolor{brightube}{rgb}{0.82, 0.62, 0.91}

\newcommand{\greyline}{\raisebox{1pt}{\tikz{\draw[-,black!40!white,solid,line width = 1.8pt](0,0) -- (5mm,0);}}}
\newcommand{\violetline}{\raisebox{1pt}{\tikz{\draw[-,violet,solid,line width = 1.8pt](0,0) -- (5mm,0);}}}

\newcommand{\eqn}[1]{\begin{align}#1\end{align}}
\newcommand{\bs}[1]{\boldsymbol{#1}}

\newcommand{\fr}[2]{\frac{#1}{#2}}

\newcommand{\avg}[1]{\langle #1 \rangle}
\newcommand{\tex}[1]{\mbox{\scriptsize{#1}}}

\newcommand{\deleted}[1]{}

\def\dd{\mathrm{d}}  
\def\kt{k_B T}

\def\bx{\bs{x}}

\def\bM{\bs{M}}

\def\bW{\bs{W}}

\def\bF{\bs{F}}

\begin{document}

\title{Metallic Microswimmers Driven up the Wall by Gravity}

\author{Quentin Brosseau$^1$, Florencio Balboa Usabiaga$^2$, Enkeleida Lushi$^3$, Yang Wu$^4$, Leif Ristroph$^1$, Michael D. Ward$^4$  Michael J. Shelley$^{1,2}$ and Jun Zhang$^{1,5,6}$}
\affiliation
{$^1$ Applied Mathematics Laboratory, Courant Institute, New York University, NY NY 10012, USA, \\
$^2$ Flatiron Institute, Simons Foundation, NY NY 10010, USA \\
$^3$ Dept. of Math. Sciences, New Jersey Institute of Technology, Newark NJ 07102, USA\\
$^4$ Dept. of Chemistry, New York University, NY NY 10012, USA\\
$^5$ Dept. of Physics, New York University, NY NY 10003, USA\\
$^6$ NYU-ECNU Institute of Physics, New York University Shanghai, Shanghai 200062, China}

\date{\today}

\begin{abstract}

As a natural and functional behavior, various microorganisms exhibit gravitaxis by orienting and swimming upwards against gravity. Swimming autophoretic nanomotors described herein, comprising bimetallic nanorods, preferentially orient upwards and swim up along a wall, when tail-heavy (i.e. when the density of one of the metals is larger than the other). Through experiment and theory, two mechanisms were identified that contribute to this gravitactic behavior. First, a buoyancy or gravitational torque acts on these rods to align them upwards. Second, hydrodynamic interactions of the rod with the inclined wall induce a fore-aft drag asymmetry on the rods that reinforces their orientation bias and promotes their upward motion.

\end{abstract}



\maketitle

As part of their survival, many microorganisms, such as the algae \textit{C. reinhardtii}, \textit{E. gracilis}, or \textit{Paramecia},
need to swim up against gravity. Such behavior is known as \textit{gravitaxis}. These swimmers, when pulled by gravity, align 
vertically due to a fore-aft drag asymmetry along their bodies that generates a hydrodynamic torque \cite{Roberts1970, Kessler1985, 
Roberts2002, Roberts2006, Roberts2010, Richter2007}. Inhomogeneous density distributions within their bodies can also lead to 
buoyancy torques and vertical alignment \cite{Mogami2001}. Once oriented vertically their propulsion allows vertical migration.
When in a group, these torques also contribute to the emergence of colonial bioconvective patterns and to the stratification of 
swimmers in the bulk \cite{Childress1975, Wolff2013, Yan2015, Ruhle2020}. Near confining walls, the dynamics of any swimmer 
is expected to change due not only to gravity but also to hydrodynamic interactions with boundaries \cite{Ruhle2018,Das2015}. Indeed, 
many microbes inhabit wet soils and other porous media where sloped boundaries are omnipresent \cite{Petroff2014, Petroff2017}. 
A natural question is whether such walls or slopes will suppress or enhance gravitaxis.

The design of artifical microswimmers can incorporate the working principles underlying organismal gravitaxis to drive, 
direct, and optimize the motion of self-propelled colloids \cite{Singh2018, Rizvi2020}.
For example, spherical polystyrene beads 
coated with a heavy metallic cap on its trailing pole and fueled by hydrogen peroxide ($\mbox{H}_2\mbox{O}_2$) swim up 
in the bulk \cite{Palacci2010, Campbell2013}.
Swimmer shape also affects trajectory, as demonstrated for L-shaped autophoretic colloids (powered by light) swimming on an inclined plane, wherein asymmetric propulsion-to-drag distribution allowed steady upslope movement plus curved motions and sedimentation \cite{tenHagen2014}.

\begin{figure}[t!]
 \centerline{\includegraphics[width=\linewidth]{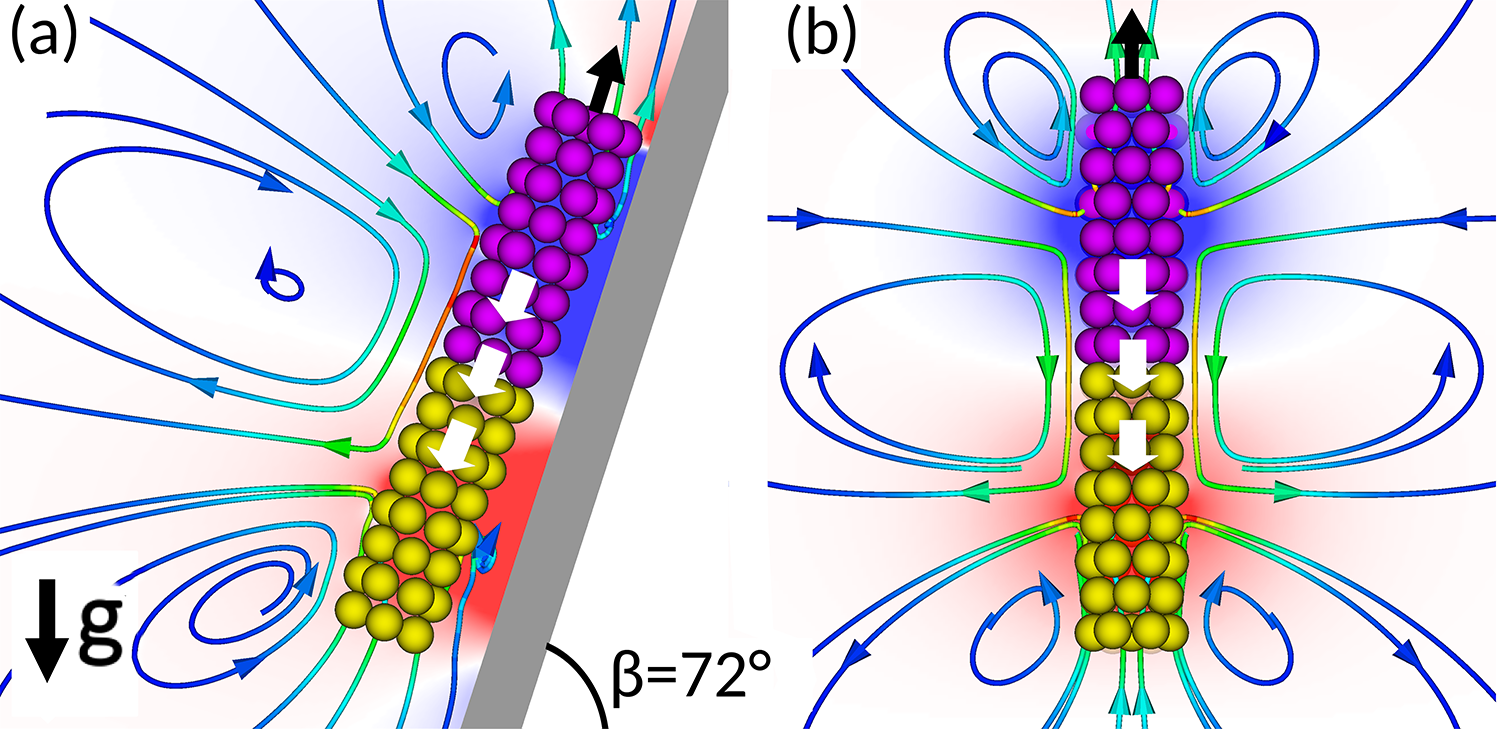}}
\caption[LoF entry]{ {\bf Climbing rod and flow fields.}
Computed flow streamlines and regions of high/low (red/blue background) pressure from a simulation of a gold-rhodium 
rod climbing a steep wall: views from the side (a) and the front of the wall (b).
In our model, the reduction and oxidation of $\mbox{H}_2\mbox{O}_2$ on the metallic segments generate an active slip layer (white 
arrows) near the bimetallic junction, propelling the rod upward. Notice that the rod has a dynamically determined head-down 
tilt with respect to the wall.}
\label{fig:flow_fields}
\end{figure}

\begin{figure}[t!]
\centerline{\includegraphics[width=\linewidth]{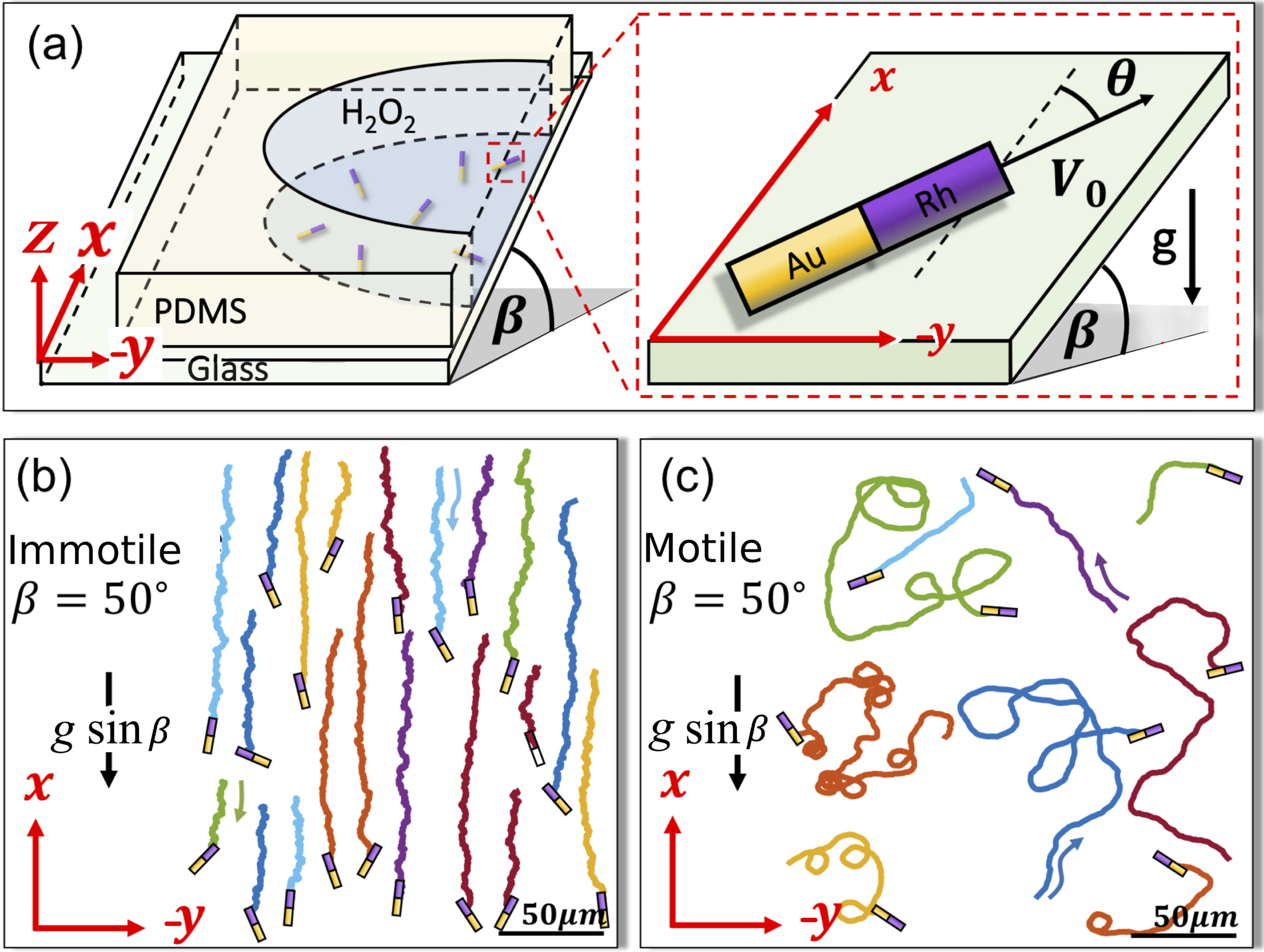}}
\caption{ {\bf Au-Rh bimetallic rods moving on an inclined wall.} (a)
  Cut-away view of our experimental setup. The rods were enclosed in a chamber containing $\mbox{H}_2\mbox{O}_2$ solution. 
  The inclination angle $\beta$ is controlled by a super-structure encasing the chamber and optical microscope (not shown). (b)
  Trajectories acquired over 2 minutes of recording for Au-Rh rods on a surface inclined $50^{\circ}$ for immotile rods (size
  exaggerated) when $\mbox{H}_2\mbox{O}_2$ is absent. (c) Motile rods (with $30\%$ $\mbox{H}_2\mbox{O}_2$) were seen to
  make random but overall upward motion against gravity, an effect more evident from statistical analysis.}
\label{fig:experimental_setup}
\end{figure}

We describe the gravitactic behavior of active bimetallic rods, combining experiments, theory and simulations to demonstrate that these heavy nanomotors can swim up inclined walls, even very steep ones; see Fig. \ref{fig:flow_fields}.
Their behavior resembles some aspects of organismal gravitaxis, as these nanomotors are tail heavy such that density inhomogeneity contributes to an upright orientation of the rods.
Direct real-time observation reveals that rods of homogeneous density sediment, i.e. do not climb, along the wall.
Surprisingly, however, these rods are subject to a gravitactic bias that slows their sedimenting speeds.
Our theoretical analysis and simulations demonstrate that the latter result can be explained by an effective fore-aft asymmetry in the hydrodynamic interaction between the rod and the nearby wall.
This additional hydrodynamic effect enhances the gravitactic behavior of rods with density inhomogeneity.

{\it Experimental Setup.} -- The bimetallic swimmers used herein were $2.5\ \si{\mu m}$ long gold-rhodium (Au-Rh) or $2.0\ \si{\mu m}$ gold-platinum (Au-Pt) rods having diameter $d \approx 0.3\ \si{\mu m}$.
The rods were synthesized by electrodeposition in anodized aluminum oxide templates according to a previously reported protocol \cite{Paxton2006, Banholtzer2009}.
The two metallic segments were either length-symmetric (1:1) Au:Rh or Au:Pt rods, or length-asymmetric (3:1) Au-Pt with long-gold and short-platinum segments. More details on the rod synthesis is  provided in the Supplementary Material \cite{SupMat}.

These rods self-propel when submersed in aqueous hydrogen peroxide ($\mbox{H}_2\mbox{O}_2$) solutions as fuel.
The fuel reduction/oxidation occurs on the Au/Pt or Au/Rh segments, creating an uneven charge distribution along the
rod.
The resulting electric field induces ionic migration in the rods' diffuse layer, creating a ``slip layer" of fluid that 
envelops the rod and is likely most pronounced at the junction between the two metals.
This fluid displacement, due to momentum conservation, results in rod movement in the opposite direction, with the rhodium 
or platinum segment leading the motion \cite{Moran2011, Moran2017}.
The geometrically symmetric gold-rhodium (Au-Rh) rods have a density asymmetry of ratio roughly
3:2 between the two segments, as $\rho_{\tex{Au}}=19.32\,
\si{g/cm^3}$ and $\rho_{\tex{Rh}}=12.41 \, \si{g/cm^3}$.
Consequently, the rod Center of Mass (CoM) sits rearwards, resulting in a tail-heavy rod.
In contrast, platinum is only slightly denser than gold,
$\rho_{\tex{Pt}}=21.45\, \si{g/cm^3}$, such that the density of Au-Pt rods is nearly balanced.
The fluid density is typically
$\rho_{\tex{f}} \approx 1.1\, \si{g/cm^3}$, depending on the amount of
$\mbox{H}_2\mbox{O}_2$ added to water.

We used a Nikon Eclipse 80i microscope mounted on a custom-made
tilting structure that permits adjustment to prescribed inclinations
from horizontal to vertical
(tilt angle $\beta \in [0,  90^{\circ}]$).
The experimental chamber was mounted on the microscope's stage 
and positioned to ensure a fixed alignment with the optics.
The chamber was a circular well with volume
approximately $1\, \si{cm^3}$, cut from a $0.5\,\si{cm}$ thick PDMS
slab and mounted on a glass slide, as illustrated in
Fig. \ref{fig:experimental_setup}a. This chamber was filled with
$\mbox{H}_2\mbox{O}_2$ solution, followed by the addition of the bimetallic rods.
The chamber was then capped with a coverslip to ensure an optically flat surface for observation and prevent fluid leakage.

The kinematic characterization of the rod swimmers was done with the chamber positioned horizontally ($\beta = 0^{\circ}$), as
the rods sediment to the bottom and move about. Their movement in the focal plane of a 40X objective lens was recorded with a camera at a rate of $25\, \mbox{frames/s}$.
Typically, the particle motion was measured for 2 minutes and their trajectories analyzed using the
{\tt  MatLab Image Processing Toolbox} and custom-written software \cite{SupMat}.
The characteristic swimming speeds under various $\mbox{H}_2\mbox{O}_2$ concentrations (between $15\%$ and $30\%$) were typically from $3$ to $8\,\si{\mu m / s}$.

\begin{figure}[t!]
\centerline{\includegraphics[width=\linewidth]{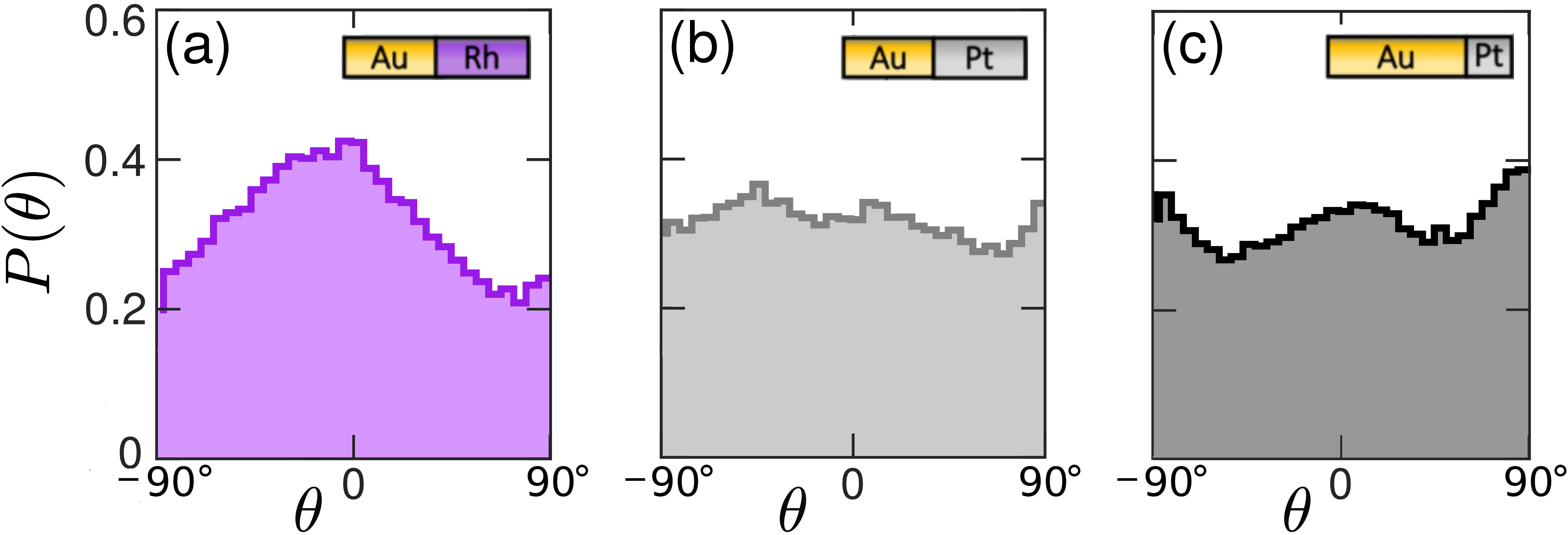}}
\caption{ {\bf Orientations of sedimenting immotile rods on
      inclined walls.} Without any $\mbox{H}_2\mbox{O}_2$, rods
  sediment due to gravity, and their angle with the $x$-axis,
  $P(\theta)$, along a wall inclined $70^{\circ}$ are shown for (a)
  Au-Rh tail-heavy rods and Au-Pt density-even rods (b)
  symmetric and (c) asymmetric with long-gold segment.}
\label{fig:distribution_theta}
\end{figure}

\vspace{0.05in}

{\it Immotile Rods on an Inclined Wall} -- In the absence of $\mbox{H}_2\mbox{O}_2$, Au-Rh and Au-Pt rods were immotile.
Since gravitational force dominates over thermal forces,
the rods, unsurprisingly, slid down in rectilinear trajectories 
(Fig. \ref{fig:experimental_setup}b).
The distribution of the angle between the rod axis and $\bs{x}$, $P(\theta)$, 
has a maximum at the vertical direction, $\theta=0$, for tail-heavy Au-Rh rods (Fig. \ref{fig:distribution_theta}a) 
and is rather flat for both types of density-balanced Au-Pt rods
(Fig. \ref{fig:distribution_theta}b, c).
In the absence of reduction/oxidation reactions (propulsion) the orientation preference
can only be linked to the density distribution of the rods.
Here, the buoyancy (geometric) center of a Au-Rh rod differs from the
CoM, giving rise to a torque that reorients
the rods.
The tail-heavy Au-Rh rods sediment with their gold ends leading and long-axis along the gravitational field.

\begin{figure}[t!]
  \centerline{\includegraphics[width=\linewidth]{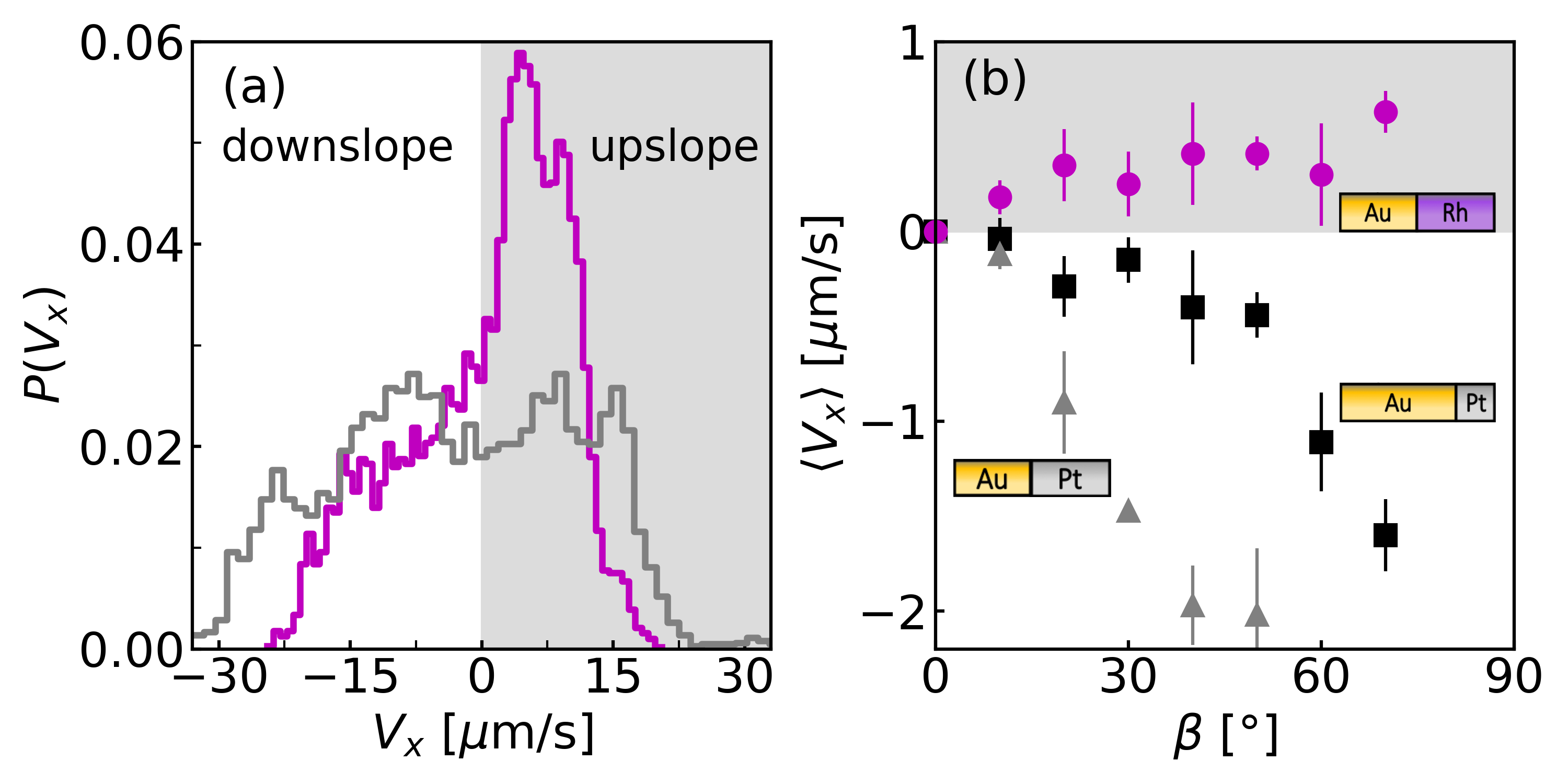}}
  \vspace{-0.1in}
\caption{ {\bf{Velocity of motile rods.}} (a) Velocity distribution of
  tail-heavy Au-Rh swimmers \protect\violetline \, and symmetric
  density-balanced, 1:1 Au-Pt swimmers \protect\greyline \, on a wall
  inclined $70^{\circ}$. Overall, the Au-Rh rods swim upslope and
  perform gravitaxis while Au-Pt rods sediment. (b) The average
  velocity along the $x$-axis $\avg{V_x}$ vs. wall inclination $\beta$
  for three rod types: Au-Rh (${\color{violet}\bullet}$),
  symmetric Au-Pt (${\color{gray}\blacktriangle}$) and long-gold Au-Pt
  (${\color{black}\blacksquare}$).  Au-Rh rods show increasing
  gravitactic ability with increased inclination, and Au-Pt rods
  sediment at different rates depending on their segmental ratios.}
\label{fig:gravi_Au_rh}
\end{figure}

{\textit{Motile Rods.}}  When submerged in an aqueous solution containing $\mbox{H}_2\mbox{O}_2$ fuel, the rods self-propel along the inclined surface/wall, as illustrated in Fig. \ref{fig:experimental_setup}c.
Their trajectories become highly
nontrivial and exhibit movement up the wall against gravity as well as sideways and downward motions.
This gravitactic behavior is made more
evident through statistical analysis of the rods motions.
As illustrated in Fig. \ref{fig:gravi_Au_rh}a, the velocity distribution $P(V_x)$ at an
inclination $\beta=70^{\circ}$ reveals that tail-heavy Au-Rh rods were
biased towards upslope swimming. Density-balanced Au-Pt rods (1:1
ratio), however, display overall downward swimming.

Fig. \ref{fig:gravi_Au_rh}b depicts the mean velocity $\avg{V_x}$
for different wall inclinations, $\beta$, and for all three swimmer
types. Tail-heavy Au-Rh rods clearly swim upslope.
This tendency increases with $\beta$, whereas
(approximately) density-balanced Au-Pt rods sediment
downslope.
Notably, symmetric Au-Pt rods sediment faster
than asymmetric 3:1 long-gold Au-Pt rods at any plane inclination
$\beta$.
The slight gain in mass in the symmetric rod due to the
longer Pt segment is not sufficient to explain its faster
sedimentation.
In the next section the role of hydrodynamic interactions between the rods and the wall in their gravitactic response and how it might control their sedimentation speed is addressed.

\begin{figure}[t!]
  \centerline{\includegraphics[width=\linewidth]{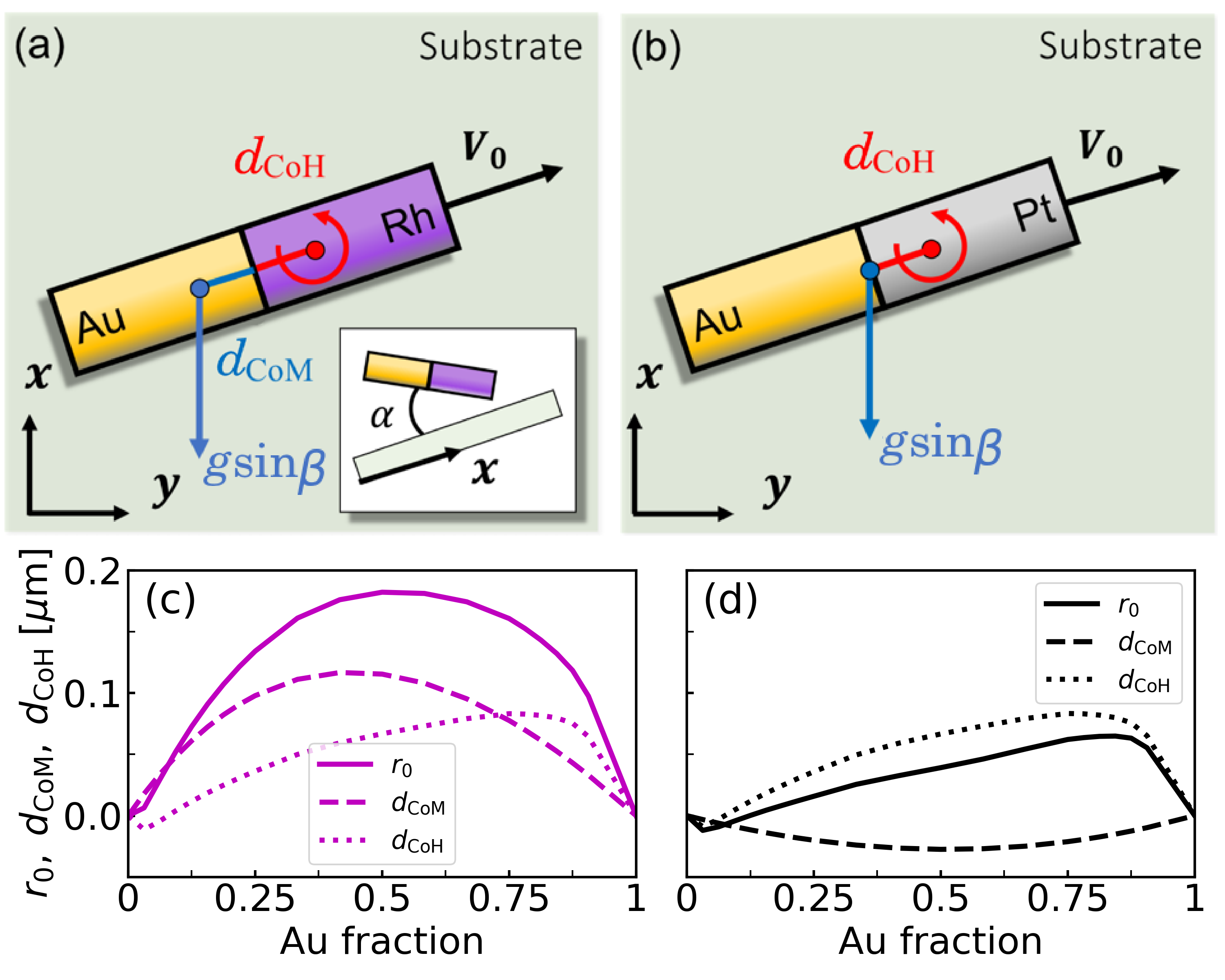}}
  \caption{{\bf{Model for gravitaxis when close to a wall.}}  Body
    forces act on the rod's center of mass (CoM) whereas rotation
    occurs around the center of hydrodynamic stress (CoH). (a) These
    tail-heavy rods experience a gravitational torque since CoM sits
    rearward and CoH headward. (b) For density-balanced
    rods, the gravity torque only appears due to the shifted CoH from
    the center. (c, d)
    Numerically obtained values for $d_{\tex{CoM}}$ and
    $d_{\tex{CoH}}$, the distances measured from the rod's center to
    the CoM and CoH, respectively, for (c) Au-Rh and (d) Au-Pt (d) rods
    with varying length of Au. Negative values of $d_{\tex{CoM}}$
    indicate that the CoM is displaced headward.}
\label{fig:CoM_CoH}
\end{figure}

\vspace{0.05in}
{\it Modeling Gravitaxis} -- Two methods were used to model gravitaxis. The first is a full, closed-up hydrodynamic description of the rods and the second 
is a simplified and zoom-out (reduced) mechanical model that predicts the rod trajectories. In the first method, each rod was modeled as a rigid body with an active slip layer centered in the bimetallic junction.
The Stokes equations were solved to determine the
surrounding flow and pressure fields in the presence of the wall, and consequently the rod orientation and swimming speed 
\cite{Usabiaga2016, Brosseau2019, SupMat}, see Fig. \ref{fig:flow_fields} .
Our second method aims to understand the observed gravitaxis of Au-Rh swimmers and the controllable sedimentation (by different segmental ratios) of Au-Pt swimmers.
The swimming rods were observed to remain close to the wall, and previous reports have revealed that immotile rods remain parallel to the wall while motile rods swim with a head-down tilt angle $\alpha$
\cite{Ren2017,Brosseau2019}; see Figs. \ref{fig:flow_fields} and \ref{fig:CoM_CoH}a inset.
The second model assumes that rod trajectories are two-dimensional, in the $xy$-plane parallel to the wall, and it describes the rod configuration by a tracking point (e.g. any fixed point on the rod)
$\bx(t) \in \mathbb{R}^2$ and the rod orientation $\theta(t)$ with respect to the $x$-axis.
The rod is now a Brownian particle with 
swimming speed $V_0$ and subject to a gravitational force $\bF$ and torque $\tau$ about the tracking point,
\eqn{
 \label{eq:model}
\left(\begin{array}{c}
  \dot{\bx} \\
    \dot{\theta}
  \end{array}\right) = \left(\begin{array}{c}
    V_0 \cos \theta \\
    V_0 \sin \theta \\
    0
  \end{array}\right) + \bM \left(\begin{array}{c}
    \bF \\
    \tau
  \end{array}\right) + \sqrt{2\kt} \bM^{1/2} \bW.}
The $3 \times 3$ mobility matrix $\bM$, calculated at the tracking
point, couples the forces and torque to the linear and angular
velocities while $\bW \in \mathbb{R}^3$ is a white noise vector that
generates the Brownian motion. The force,
$\bF = -m g \bs{e}_x \sin \beta$, is proportional to $m$ the rod excess of mass over the displaced fluid, acceleration due to
gravity $g$ and increases with the wall inclination $\beta$. The
gravitational reorienting torque $\tau$ (normal to the $xy$-plane)
has magnitude $r_0 m g \sin \beta \cos \alpha \sin \theta$, where
$r_0$ is the lever arm, i.e. the distance between the tracking point
and the rod CoM.

Analyzing \eqref{eq:model} is generally difficult because the rod translational and rotational dynamics are coupled.
However, the orientation equation can be decoupled from the translation if the tracking point is chosen to be the center of rotation.
The center of rotation is defined to be a pivot point about which an applied torque generates only rotation and not translation, whereas 
a net body force generates only translation and not rotation.
Such a pivot point is known to exist for two dimensional motion \cite{Brenner_book, Bernal1980, Delong2015b}.
We denote this pivot point as the
\emph{Center of Hydrodynamic stress} (CoH).
Using the CoH as the tracking point the orientation equation simplifies to
\eqn{
\label{eq:orientation}
\fr{\dd \theta}{\dd t} = M_{\omega \tau} \tau + \sqrt{2\kt} M_{\omega \tau}^{1/2} W_{\theta}.
}

For a rod in the bulk, far from any walls, the CoH is located at its geometric center.
When near to a wall the CoH location may shift.
For swimmers with a head-down tilt ($\alpha > 0$) the increased
resistance near the front displaces the CoH headward from the
geometric center of the rod \cite{SupMat}.
Therefore, the lever arm at which a body force exerts a gravitational torque can be decomposed
into two contributions, $r_0=d_{\tex{CoM}}+d_{\tex{CoH}}$,
i.e. distances measured from the rod's center to the CoM and CoH,
respectively; see Figs. \ref{fig:CoM_CoH}a, b.  Overall, the larger
the lever arm $r_0$, the larger the reorienting torque. This increased torque can come to dominate disorienting thermal fluctuations.
Thus, once oriented upwards by the gravitational torque, a rod swimmer may move upwards gravitactically.

A sizable level arm $r_0$ can be achieved using metals with density
contrast (e.g. in the Au-Rh case, $d_{\tex{CoM}}$ sits rearwards) or
with different segmental lengths of the two metals (e.g. in the Au-Pt
3:1 case, $d_{\tex{CoH}}$ is shifted headwards) \cite{SupMat}.
For the Au-Rh rods, the distance from the rod's center to its CoM, $d_{\tex{CoM}}$, is maximized for approximately symmetric rods, i.e. $L_{\tex{Au}} \approx L / 2$, see Figs. \ref{fig:CoM_CoH}c.

The $d_{\tex{CoH}}$ can be increased by moving the metal junction, and thus the location of the slip layer, headward.
This fluid layer, which propels the rods, creates a pressure field that tilts the rods \cite{Brosseau2019}. Such head-down tilt (angle $\alpha$ up to
moderate values, $\alpha \le 10^{\circ}$ \cite{SupMat}) makes the leading portion of the rod closer to the solid wall than the trailing
portion, see Fig. \ref{fig:flow_fields}. The resulting resistance difference,
higher near the head but lower at the tail, shifts the CoH headward and thus increases $d_{\tex{CoH}}$.
Therefore, the location of the junction largely determines the position of the CoH. Figs. \ref{fig:CoM_CoH}c, d illustrate the values (dotted curves) of $d_{\tex{CoH}}$, as functions of the position where two metals join, obtained with our full hydrodynamic model \cite{SupMat}.
Combining both contributions to the lever arm, $r_0$ (solid curves in
Fig. \ref{fig:CoM_CoH}c, d), the model predicts that the gravitactic effect for Au-Rh rods will be maximized for length-symmetric swimmers
while for Au-Pt rods will be maximized for rods with length asymmetric long gold segments.
This is consistent with our experimental results shown in
Fig. \ref{fig:gravi_Au_rh}, and some of these predictions are validated in the next section.

\begin{figure}[t!]
  \centerline{\includegraphics[width=\linewidth]{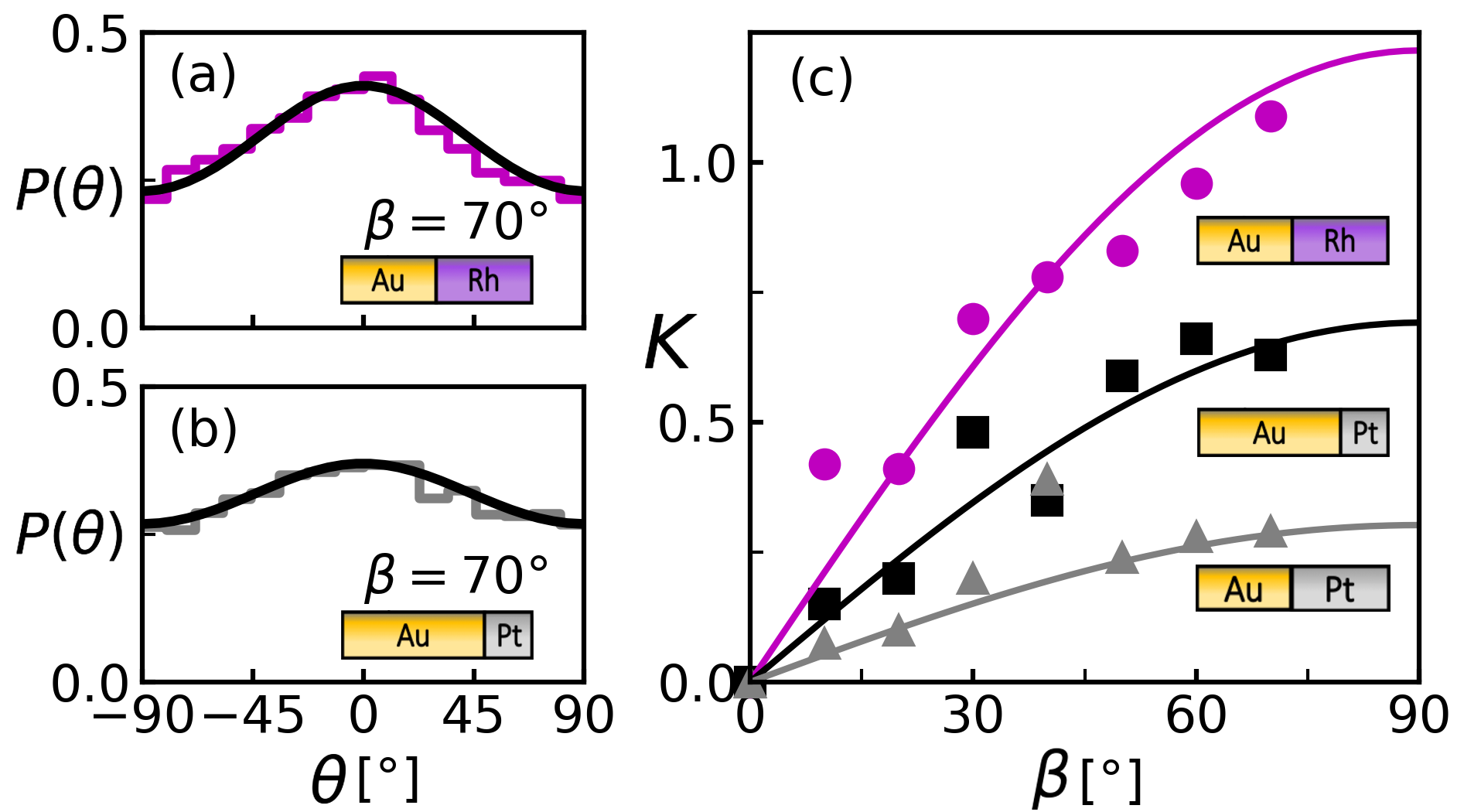}}
\caption{ {\bf{Experiment vs. model.}} Experimental orientation
  distributions for motile (a) tail-heavy Au-Rh rods, and (b)
  asymmetric density-even Au-Pt rods, show peaks at
  $\theta=0^{\circ}$. Fitting the data by Eq. \eqref{eq:Ptheta} (solid
  curves) we obtained $K$. (c) The extracted $K$ values are plotted
  vs. tilt angle $\beta$.  Values of $r_0$, further extracted by
  fitting $K\sim r_0 \sin\beta$ (solid lines), show gravitational
  torques act at lengths greater than $d_{\tex{CoM}}$ due to the
  shifted CoH in all 3 cases [tail-heavy
    (${\color{violet}{\bullet}}$), density balanced asymmetric
    (${\color{black}{\blacksquare}}$) and density balanced symmetric
    rods (${\color{gray}{\blacktriangle}}$)].}
\label{fig:active_fit_model}
\end{figure}

\vspace{0.05in}
{\it Quantifying the Lever Arm} -- To test the coupled effects of gravity and hydrodynamic
interactions with the wall, we examine the orientation of motile rods.
The equilibrium distribution of the angle between the rod and the $x$-axis predicted by the mechanical model is (from Eq.
 \eqref{eq:orientation})
\footnote{The PDF of $\theta$ predicted by the Eq. \eqref{eq:orientation} is
  $P'(\theta) = \exp(K\cos \theta) / (2\pi I_0(K))$.  However, the
  experiments measure the angle formed by the rod axis with the
  $x$-axis and do not distinguish the orientation $\theta$ from
  $\theta'=\theta - \pi$.  The formula given in the main text,
  $P(\theta) = P'(\theta) + P'(\theta - \pi)$, is the one used to fit
  the experiments.},
\eqn{
  \label{eq:Ptheta}
  P(\theta) = \fr{e^{K \cos \theta} + e^{-K \cos \theta}}{2\pi I_0(K)},}
 where $I_0(K)$ is the modified Bessel function of order
zero and $K = r_0 m g \sin \beta \cos \alpha / \kt$ is
the ratio between the gravitational torque and the thermal
energy, which tends to randomize the rod orientation. Upward
swimming is possible when $K$ is larger than the ratio between the
sedimentation velocity and the intrinsic swimming speed $V_0$  \cite{SupMat}.
The experimental results and theoretical curve fit are depicted in Fig. \ref{fig:active_fit_model}a and b.
The peaks appearing at $\theta=0$ for both Au-Rh and Au-Pt rods are consistent with the existence of a lever arm $d_{\tex{CoH}}$ predicted by the mechanical model.

From $P(\theta)$ we extracted the parameter $K$ that best fits the experimental results using Eq. \eqref{eq:Ptheta}; the values of $K$ versus the wall inclination are shown in Fig. \ref{fig:active_fit_model}c.
As expected  from the model, $K$ is indeed proportional to $\sin\beta$.
The results demonstrate that the overall torque is higher for Au-Rh rods, for which $d_{\tex{CoM}}$ is non-negligible.
The values of the lever arm $r_0$ can be extracted by fitting the values of $K$ to the mechanical model prediction.
For Au-Rh rods the fit yields $r_0= 0.19\, \si{\mu m}$, corresponding to $d_{\tex{CoH}}=0.05\, \si{\mu m}$, which is ahead of the rod's
midpoint because its $d_{\tex{CoM}}=0.14\, \si{\mu m}$.
Here, hydrodynamic effect accounts for about $25\%$ of the torque felt by tail-heavy rods.
 
The Au-Pt rods were slightly head-heavy as platinum is denser than gold.
In the cases of symmetric 1:1 Au:Pt and front-actuated 3:1 Au:Pt rods, $d_{\tex{CoM}}$ is $-0.026$ and $-0.02\ \si{\mu  m}$, respectively.
This contribution is insufficient to produce a bias in the rod
orientation.
The experimental data suggest a torque larger than the
one created only by the density mismatch. A fit of the experimental
results reveals that the distance of the CoH to the geometric center
is larger for asymmetric rods ($d_{\tex{CoH}} = 0.14\, \si{\mu m}$)
than for the symmetric ones ($d_{\tex{CoH}}=0.076\, \si{\mu m}$).
This arm length difference generates the distinct sedimentation speeds of our two Au-Pt rod types.

\vspace{0.05in}
{\it  Conclusion} -- These results demonstrate gravitaxis using density unbalanced nanomotors fueled with $\mbox{H}_2\mbox{O}_2$.
These ``cliff climbers'', which are about 15-20 times heavier than the surrounding fluid, move up steep walls.
Interestingly, it is the gravitational pull that orients these
tail-heavy rods and allows gravitaxis.
Moreover, the emergent hydrodynamic effect when rods interact with the sloped walls \cite{Spagnolie2012, Brosseau2019} enhances the effect.
Such enhancement can be used to control the sedimentation speed of
falling rods and promote gravitaxis.

The microswimmer behavior clearly reveals that an imbalance in density of the two metals results in a reorienting gravitational torque, due to the shift of its center of mass (CoM).
Additionally, the shift of the center of rotation reveals the importance of the hydrodynamic interactions.
Both effects take place and contribute to successful gravitaxis.
The lessons learnt here in artificial systems might overlap with behaviors observed in organisms.
Specifically, heterogeneous density distribution and hydrodynamic effects may assist swimming bacteria and other microorganisms to perform gravitaxis for their survival.
Indeed, it is well established that many microswimmers are attracted
to walls by hydrodynamic interactions \cite{Berke2008, Takagi2014,
  Contino2015, Lushi2017}.
Once near a wall, if the swimmer is denser than the fluid, as usually happens, it may be oriented upwards by a gravity-induced torque and then climb up the wall.
This phenomenon could affect the distribution of microorganisms in porous soils \cite{Kuznetsov2001, Petroff2014, Petroff2017}.
It would be interesting to examine if any microorganism capitalizes on this mechanism to control vertical migration in a complex environment.

This work was supported primarily by the MRSEC Program of the National Science Foundation under Award DMR-1420073, and also by NSF Grants DMS-RTG-1646339, DMS-1463962 and DMS-1620331.

Q.B. and F.B.U. contributed equally to this work.

\bibliography{Biblio_gravi.bib}

\end{document}


\title{Metallic Microswimmers Driven up the Wall by Gravity}

\author{Quentin Brosseau$^1$, Florencio Balboa Usabiaga$^2$, Enkeleida Lushi$^3$, Yang Wu$^4$, Leif Ristroph$^1$, Michael D. Ward$^4$  Michael J. Shelley$^{1,2}$ and Jun Zhang$^{1,5,6}$}
\affiliation
{$^1$ Applied Mathematics Laboratory, Courant Institute, New York University, NY NY 10012, USA, \\
$^2$ Flatiron Institute, Simons Foundation, NY NY 10010, USA \\
$^3$ Dept. of Math. Sciences, New Jersey Institute of Technology, Newark NJ 07102, USA\\
$^4$ Dept. of Chemistry, New York University, NY NY 10012, USA\\
$^5$ Dept. of Physics, New York University, NY NY 10003, USA\\
$^6$ NYU-ECNU Institute of Physics, New York University Shanghai, Shanghai 200062, China}

\date{\today}

\maketitle

\section{Experimental details (sample preparation)} 
The nanorods are synthesized by a templating method on Anodic Aluminium Oxide (AAO) membranes (Whatman Anodisc™ 47) with a typical pore diameter of $0.3\, \si{\mu m}$.  Prior to the electrodeposition, one side of the AAO membrane is sealed by thermo-evaporation of a
$150\, \si{nm}$ thick layer of silver (BAL-TEC MCS 010 Multi Control System).

The electrodeposition is made in three steps using a three-electrodes method:
\begin{itemize}
\item{A layer of silver is deposited  at $-1\, \si{V}$ from an aqueous solution of silver cyanide (0.0186 M, AgCN, Thermo Fisher Scientific Inc.), potassium cyanide (0.1233 M, KCN, Thermo Fisher Scientific Inc.) and potassium pyrophosphate (0.0304 M, $\mbox{K}_4\mbox{P}_2\mbox{O}_7$, Sigma-Aldrich, Co. LLC) to prevent leakages.}

\item{A layer of gold is deposited at $-0.92 \si{V}$ from a commercial plating solution (OROTEMP 24 RTU Rack from TECHNIC INC).}

\item{In the case of Au-Rh symmetric nanorods, a layer of rhodium is deposited at $-0.4\, \si{V}$ from a commercial plating solution (Techni Rhodium RTU from TECHNIC INC). In this case, the deposition charges  of gold $C_{\tex{Au}}$ and Rhodium $C_{\tex{Rh}}$ are $16\,\si{C}$ and $68\,\si{C}$, respectively.

In the case of Au-Pt nanorods, a layer of platinum is deposited at $-0.4\,\si{V}$ from an aqueous solution of  ammonium hexachloroplatinate (IV) (0.010 M, (NH$_3$)$_2$PtCl$_6$, Alfa Aesar) and sodium phosphate dibasic dihydrate (0.020 M,$\mbox{Na}_2\mbox{HPO}_4$, Sigma-Aldrich, Co. LLC). The deposition charges  of gold and platinum are $C_{\tex{Au}}=7.2\,\si{C}$ and platinum $C_{\tex{Pt}}=26\,\si{C}$ for symmetric rods  and $C_{\tex{Au}}=24\,\si{C}$ and  $C_{\tex{Pt}}=9\,\si{C}$ for long gold segment rods.}
\end{itemize}

The silver layer is etched away in a solution of $\mbox{HNO}_3$ (1 M), and  the membrane is dissolved in a NaOH solution (5 M).
The resulting suspension with nanorods is purified through a repeated centrifugation/dilution process.

\subsection{Rods' parameters}
Table \ref{tab:swim_props} shows the geometry of the rods, their diffusion coefficients ($D_t$ and $D_r$) and their swimming speeds ($V_0$) at different $\mbox{H}_2\mbox{O}_2$ concentration.

\begin{table}
\caption{\label{tab:swim_props} Table of geometrical and physical properties of symmetric Au-Rh nanorods and the two Au-Pt nanorod types with gold and platinum fractions $(l_{\tex{Au}}:l_{\tex{Pt}})$.}
\begin{ruledtabular}
\begin{tabular}{lcrrrrr}
 Batch & $D_t \,[\si{\mu m^2/s}]$ &$ D_r \,[\si{1/s}]$ & $V_{0}\, [\si{\mu m /s}]$& H$_2$O$_2\%$ \\
\hline
Au-Rh(1:1) $L=2.5\,\si{\mu m}$ & 0.3& 1.04 & 0 &0 \\
Au-Rh(1:1) $L=2.5\,\si{\mu m}$ & 0.3& 1.04  & $2.7  \pm 0.3$ &10 \\
Au-Rh(1:1) $L=2.5\,\si{\mu m}$ & 0.3& 1.04  & $4.5 \pm 0.5$ &15\\
Au-Rh(1:1) $L=2.5\,\si{\mu m}$ & 0.3& 1.04 & $8.0 \pm 1.0$  &30\\
Au-Pt(1:1) $L=2\,\si{\mu m}$ & 0.25& 0.6 & $6.1 \pm 0.8$  &15\\
Au-Pt(3:1) $L=2\,\si{\mu m}$&0.25 & 0.6 & $6.5 \pm 0.7$ & 20\\
\end{tabular}
\end{ruledtabular}
\end{table}

\clearpage

\section{Center of Hydrodynamic Stress (CoH)}
\label{sec:CoH}
The choice of a tracking point to describe the orientation of a body
cannot affect the dynamics of the system (of course!) but it affects the structure of the equations of motion. 
In systems with inertia, it is natural to choose the center of mass (CoM) as the tracking point because the translational and rotational contributions to the kinetic energy decouple. 
For particles immersed in a Stokes flow, where inertia does not play a role, other choices are more convenient. 
This issue was explored by Brenner in the 1960s \cite{Brenner_book} and later expanded and clarified by Garc{\'i}a Bernal and Garc{\'i}a de la Torre \cite{Bernal1980}. 
We reproduce here their principal results for completeness. 

\begin{figure}
  \includegraphics[width=0.39 \linewidth]{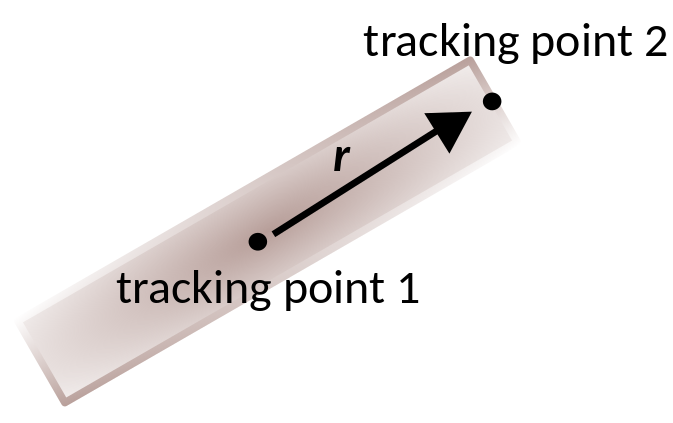}
  \caption{Sketch of a rigid body with two tracking points.}
\label{fig:sketch_tracking_points}
\end{figure}

Consider a rigid body immersed in a three dimensional Stokes flow.
Its dynamics can be described by the linear and angular velocity about a tracking point $1$.
The linear system that relates the force and torque ($\bF_1$ and $\btau_1$) with the linear and angular velocities ($\bV_1$ and $\bomega_1$) is
\eqn{
\label{eq:linearSystem}
\left(\begin{array}{c}
\bV_1 \\
\bomega_1 
\end{array}\right) = 
\left(\begin{array}{cc}
 \bM_{VF, 1} & \bM_{\omega F, 1}^T \\
 \bM_{\omega F, 1} & \bM_{\omega \tau, 1}
\end{array}\right) \left(\begin{array}{c}
\bF_1 \\
\btau_1
\end{array}\right),
}
where the $3 \times 3$ mobility components $\bM_{VF,1}$ etc. depend on the tracking point chosen to describe the motion as indicated by the subindex $1$. 
The force, torque and velocities defined at a second tracking point are
\eqn{
\label{eq:transformation_force}
\bF_2 = \bF_1, \;\;\; \btau_2 = \btau_1 - \br \times \bF_1, \\
\label{eq:transformation_velocity}
\bV_2 = \bV_1 + \bomega_1 \times \br,\;\;\; \bomega_2 = \bomega_1,
}
where $\br$ is the vector that goes from the first to the second tracking point, see Fig. \ref{fig:sketch_tracking_points}.
We can use \eqref{eq:linearSystem}-\eqref{eq:transformation_velocity} to show that the mobility components transform between tracking points like \cite{Bernal1980}
\eqn{
\bM_{\omega \tau, 2} &= \bM_{\omega \tau, 1}, \\
\bM_{\omega F, 2} &= \bM_{\omega F, 1} + \bM_{\omega \tau, 1} \times \br \\
\bM_{V F, 2} &= \bM_{V F, 1} - \br \times \pare{\bM_{\omega \tau, 1} \times \br} + \bM_{\omega F, 1}^T \times \br - \br \times \bM_{\omega F, 1},
}
where the cross product between a $3\times 3$ matrix and a vector is defined, using the Levi-Civita symbol, as 
$(\bM \times \br)_{ij} = M_{ik} \epsilon_{jkl} r_l$ and
$\br \times \bM = - \bM \times \br$.

For any body shape there is a special tracking point where the coupling matrix $\bM_{\omega F}$ is symmetric. This point is called in the literature the \emph{Center of Mobility} \cite{Bernal1980}. The location of the center of mobility with respect to an arbitrary tracking point can be found by solving for $i\ne j$ the linear system \cite{Delong2015b}
\eqn{
\label{eq:center_mobility}
\pare{\epsilon_{ikl} \pare{\bM_{\omega \tau}}_{jk} - \epsilon_{jkl}\pare{\bM_{\omega \tau}}_{ik}} r_l = \pare{\bM_{\omega F}}_{ij} - \pare{\bM_{\omega F}}_{ji},
}
where the mobility components are calculated at the original tracking point. For bodies of enough symmetry
(e.g. axisymmetric bodies), the coupling matrix 
$\bM_{\omega F}$ vanishes at the center of mobility.
In such cases, the center of mobility is also called the \emph{Center of Hydrodynamic stress} (CoH). Therefore the CoH can be found from \eqref{eq:center_mobility} if it exists. 

For two dimensional systems (or three dimensional particles constrained to move in the $xy$ plane), the CoH always exists and it corresponds to the point where a torque applied out of the plane does not generate translations. We can compute its location respect an arbitrary tracking point with \cite{Delong2015b}
\eqn{
\br = \pare{\fr{M_{\omega_z F_y}}{M_{\omega_z \tau_z}}, \fr{M_{\omega_z F_x}}{M_{\omega_z \tau_z}}}.
}
As discussed in the main text, we use this tracking point to uncouple the rotational equation of motion from translations.

\section{Effect of the swimming speed}
In this section we estimate the critical swimming speed that allows upward movement (i.e. $\avg{V_x}>0$). When the reorienting torque is very large (i.e. $K \gg 1$) the rods are aligned with the $x$-axis and the critical swimming speed coincides with the speed of sedimentation along the wall
$V_{0c} = \mu_{\parallel} m g\sin \beta$, here $\mu_{\parallel}$ is the tangential mobility of the rod.
For weak reorienting torques, $K \lessapprox 1$, like in our experiments the critical swimming speed will be larger as the rods are not always aligned with the $x$-axis.
We use our mechanical model to estimate $V_{0c}$ in this regime.
The linear velocity of the Center of Hydrodynamic stress (CoH) is (from eq. [1] in the main text)
\eqn{
  \label{eq:u}
  \bV = \left(\begin{array}{c}
    V_0 \cos \theta \\
    V_0 \sin \theta
  \end{array} \right) + \bM_{VF} \bF + \mbox{[noise terms]},
}
note that the torque does not appear explicitly because we use the CoH as the tracking point. The mobility depends on the angle $\theta$ between the rod and the $x$-axis
\eqn{
  \label{eq:M}
  \bM_{VF} = \left(\begin{array}{cc}
    \cos \theta & -\sin \theta \\
    \sin \theta & \cos \theta
  \end{array} \right)
  \left(\begin{array}{cc}
    \mu_{\parallel} & 0 \\
    0 & \mu_{\perp}
  \end{array} \right)
  \left(\begin{array}{cc}
    \cos \theta & -\sin \theta \\
    \sin \theta & \cos \theta
  \end{array} \right)^T,
}
where $\mu_{\parallel}$ and $\mu_{\perp}$ are the parallel and perpendicular mobilities (for a slender body
$\mu_{\parallel} = 2 \mu_{\perp}$).
Therefore, the velocity along the $x$-axis is
\eqn{
  \label{eq:ux}
  V_x = V_0 \cos \theta + \mu_{\perp} F_x + (\mu_{\parallel}-\mu_{\perp})F_x \cos^2 \theta +
  \mbox{[noise terms]},
}
and after integrating over orientations we get
\eqn{
\avg{V_x} = \fr{1}{2\pi}\int_{-\pi}^{\pi} V_x P(\theta)\dd \theta = V_0 \fr{I_1(K)}{I_0(K)} + \fr{1}{2}\corchete{(\mu_{\parallel}+\mu_{\perp}) + (\mu_{\parallel}-\mu_{\perp}) \fr{I_2(K)}{I_0(K)}} F_x,
}
where $I_n(x)$ are modified Bessel functions of the first kind and
we used the angle distribution,
$P(\theta) = \exp(K\cos\theta)/(2\pi I_0(K))$, obtained from the eq. [2] in the main text. To first order in $K$ 
\eqn{
  \avg{V_x} = \fr{V_0 K}{2} +  \fr{\mu_{\parallel}+\mu_{\perp}}{2} F_x.
}
Upward swimming ($\avg{V_x} > 0$) is possible when
$K > -(\mu_{\parallel}+\mu_{\perp})F_x / V_0$, i.e. when $K$ is larger than the ratio between the sedimentation velocity and the intrinsic swimming speed.
After substituting the values of the gravitational force, $F_x=-mg\sin \beta$, and 
$K = r_0 m g \sin \beta \cos \alpha / \kt$, we obtain the critical swimming speed for weak reorienting torques 
\eqn{
  V_{0c} = \fr{(\mu_{\parallel}+\mu_{\perp}) \kt}{r_0 \cos \alpha}.
}
Interestingly, in this regime the critical swimming speed is independent of the particle mass and the inclination of the wall, as long as $mg \sin \beta > 0$, because the gravitational pulling force contributes both to reorient the particle upwards and to pull it downwards.
The length of the lever arm, $r_0$, is critical.

\begin{figure}
  \includegraphics[width=0.49 \linewidth]{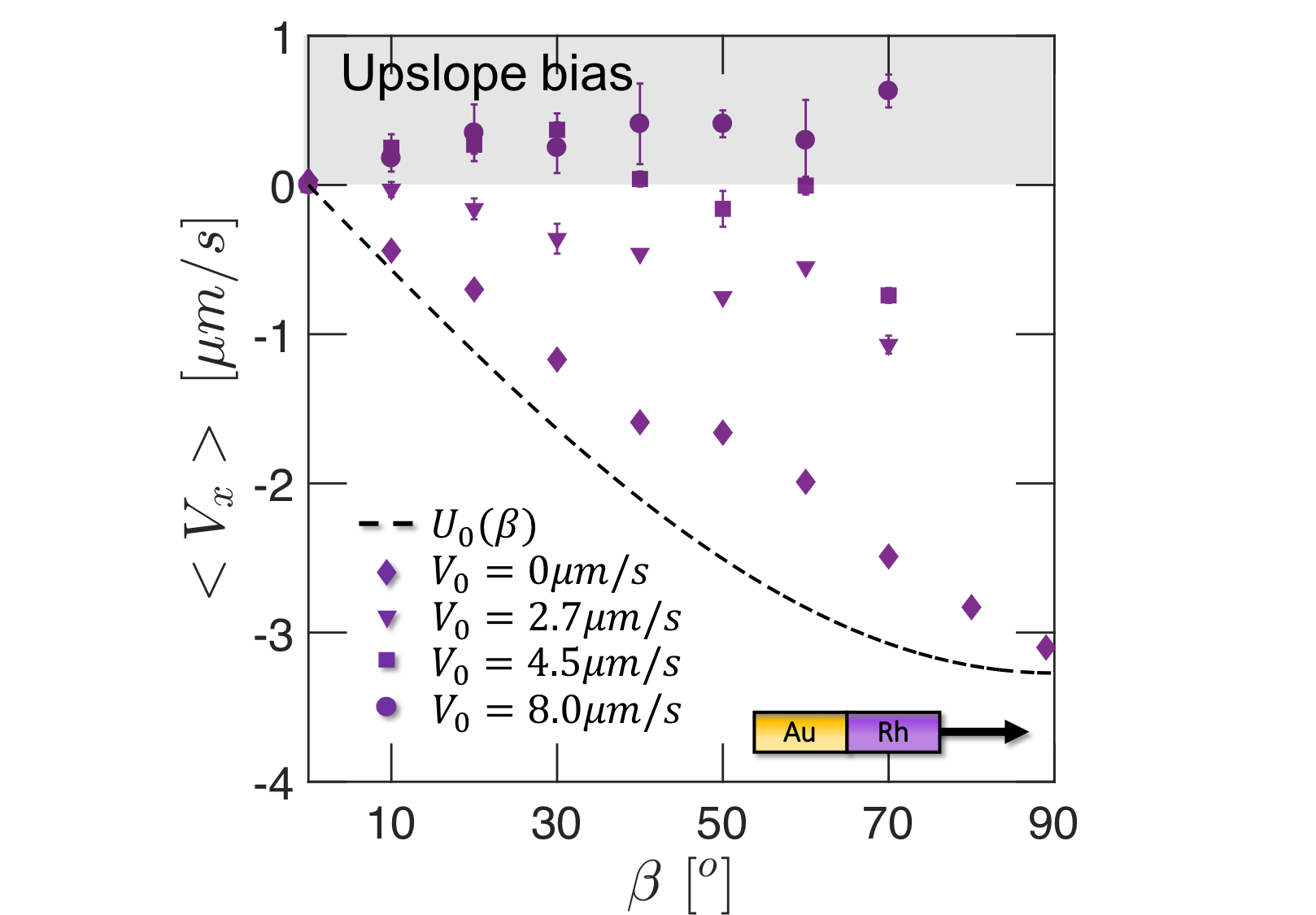} 
  \caption{Experimental results (symbols) of gravitaxis of bottom-heavy rods for several swimming speeds controlled with the $\mbox{H}_2\mbox{O}_2$ concentration. The dashed line represents the sedimentation velocity estimation
$U_0(\beta) = -\mu_{\parallel}m g \sin \beta$.}
\label{fig:gravitaxis}
\end{figure}

Using the mobility approximation of a cylinder in bulk ($\mu_{\parallel}=(\log(L/r)-0.72)/(2\pi\eta L)$ and
$\mu_{\perp} = \mu_{\parallel}/2$ \cite{Brenner_book}) we estimate a critical swimming speed of around
$4\, \si{\mu m / s}$ for our Au-Rh particles.
Indeed, Fig. \ref{fig:gravitaxis} shows that slow rods fall for all wall inclinations.
Faster rods with  $V_0 = 4.5\, \si{\mu m/s}$ show upslope motion (shaded area) for moderate values of $\beta$. For higher swimming speeds, upslope motion is visible for all inclinations.

\section{Hydrodynamic numerical model}
\label{sec:model}

We provide here some additional results obtained from numerical simulations to support our claims.
We model the rods as rigid particles immersed in a Stokes flow. We solve the Stokes equation to determine the equilibrium position of active rods with respect to the wall. 
In some simulations we include thermal fluctuations (Brownian motion)
while solving the hydrodynamic problem to compute the average velocity up the wall and $K$, the ratio between the gravitational torque and the thermal energy.
In all simulation we use the Rigid multiblob method described in Refs. \cite{Delong2015b, Usabiaga2016, Sprinkle2017}.
Details of the modeling are given in Ref. \cite{Brosseau2019}; the main difference is the length of the active slip on the rods' surface as explained next.

In the experiments a electrochemical reaction creates an active slip near the rod surface \cite{Moran2011, Moran2017}. 
In our numerical modeling instead of solving the chemical reaction we assume that part of the rod's surface is cover by an active slip of constant magnitude, $\wtil{u}_s = 30\, \si{\mu m / s}$, parallel to the axis of the rod.
In a previous publication we assumed that the active slip covered half the rod's length and that it was centered at the metal-metal interface (i.e. at the gold rhodium or gold platinum interface) \cite{Brosseau2019}.
This modeling choice was motivated by the redox nature of the chemical reaction \cite{Moran2011} and it was sufficient to explain the rheotaxis of phoretic rods in shear flows \cite{Brosseau2019}.
Here we assume again that the active slip is centered at the metal-metal junction but allow the active slip to cover less than half of the rod.
We assume that the length of active slip is
\eqn{
L_s = \left\{\begin{array}{cc}
2 L_{\tex{Au}} & \mbox{if } L_{\tex{Au}} \le L/4, \\
L/2 & \mbox{if } L/4 \le L_{\tex{Au}} \le 3L/4 \\
L - 2L_{\tex{Au}} & \mbox{if } L_{\tex{Au}} \ge 3L/4,
\end{array}\right. 
}
where $L$ is the length of the rod and $L_{\tex{Au}}$ the length of the gold segment.

\begin{figure} 
  \includegraphics[width=0.95 \columnwidth]{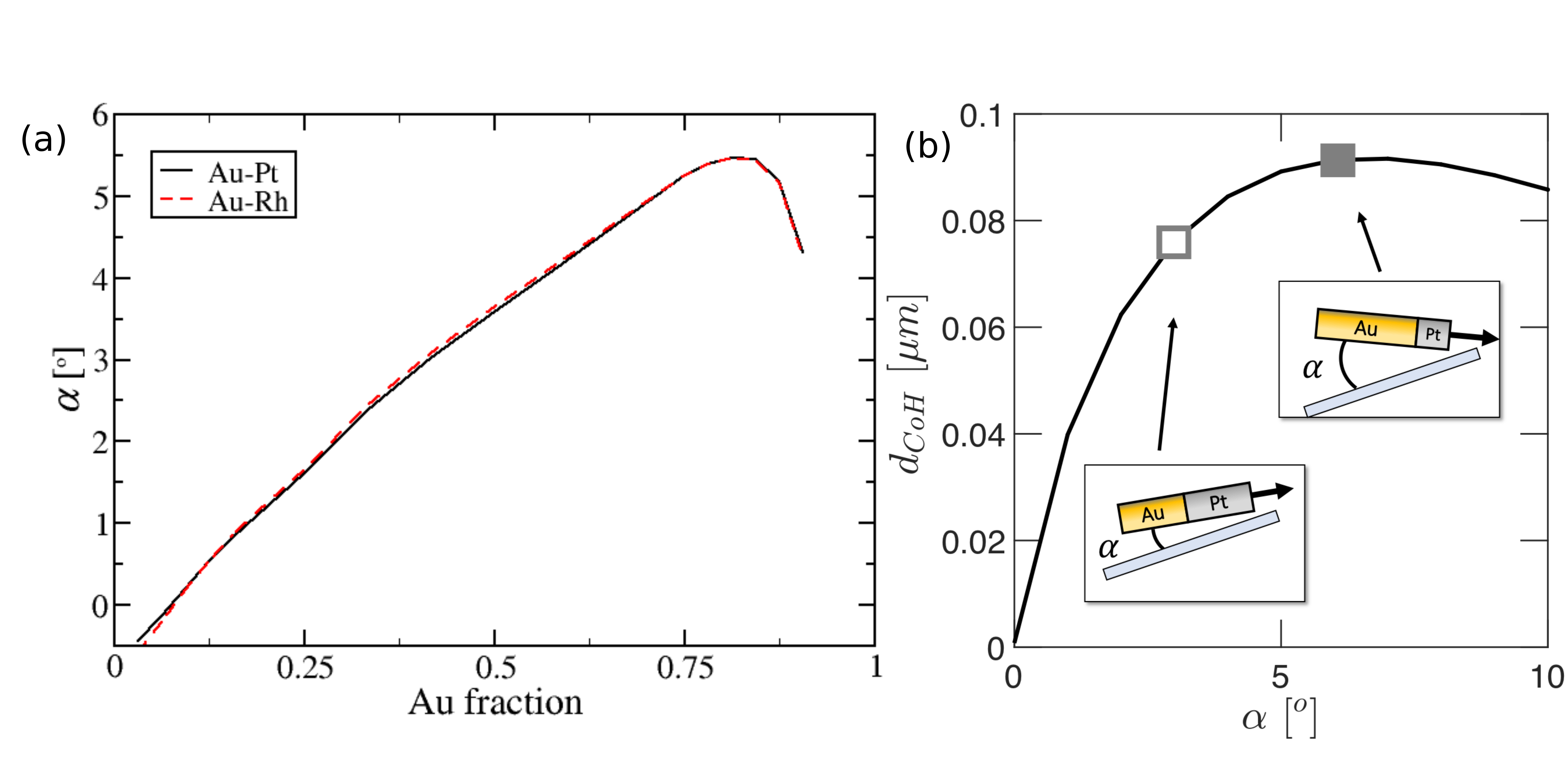}
\caption{Results from deterministic simulations.
{\bf (a)} tilt angle $\alpha$ with the wall for rods with different gold fractions. All rods have length $L=2\, \si{\mu m}$.
{\bf (b)} Distance between the CoH and the rod's geometric center assuming that the rod's head is at a height $h=0.2\, \si{\mu m}$ from the wall.}
\label{fig:angle_d_CoH}
\end{figure}

We show the tilt angle $\alpha$ of rods towards the wall in the left panel of figure \ref{fig:angle_d_CoH}. Rods with long gold segments
tilt more. This is consistent with our previous investigation about the dynamic of phoretic swimmers in shear flows \cite{Brosseau2019}.
Note that the tilt angle is controlled by the location of the active slip along the rod and that the density difference between Au-Rh and Au-Pt plays a minimal role.
We show the computed distance between the CoH and the rod's center in Fig. \ref{fig:angle_d_CoH} right. The physical interpretation of these results is that the higher drag near the front of the rod displaces the center of rotation forward. 
Note that in the limit where a rod is pinned to the wall the
anchor point will act as the center of rotation.

\subsection{Additional numerical results}
\label{sec:numerical_results}
\begin{figure}
  \includegraphics[width=0.45 \columnwidth]{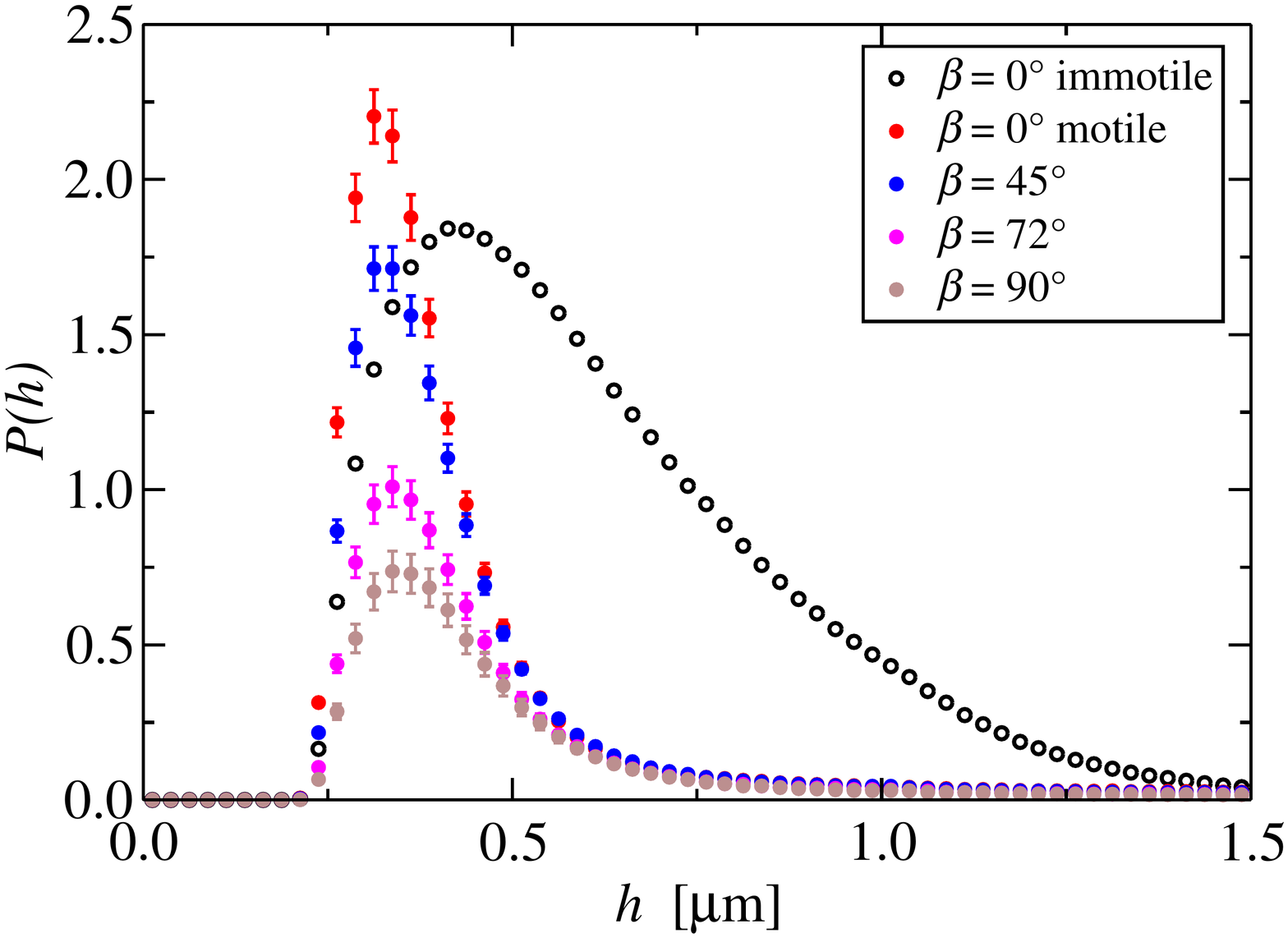}
  \caption{Equilibrium distribution of the distance between 
    $2\, \si{\mu m}$ long rods and the wall obtained from Brownian simulations and different wall inclinations $\beta$. Results for motile rods with swimming speed
    $V_0=10.6\, \si{\mu m/s}$ (full symbols) and immotile rods (open symbols).}
  \label{fig:distance_wall}
\end{figure}

Here we present some additional results obtained from simulations. 
First, Fig. \ref{fig:distance_wall} shows the equilibrium distribution, $P(h)$, of the distance between the bottom-heavy rods and the wall.  The osmotic flow around active rods draws them closer the wall. The distributions show that most of the rods lay at a distance comparable to a rod diameter
($d=0.3\, \si{\mu m}$). The mean distance varies marginally between rods on horizontal ($\beta=0^{\circ}$) and vertical ($\beta=90^{\circ}$) planes. 
    
\begin{figure}
  \includegraphics[width=0.49 \columnwidth]{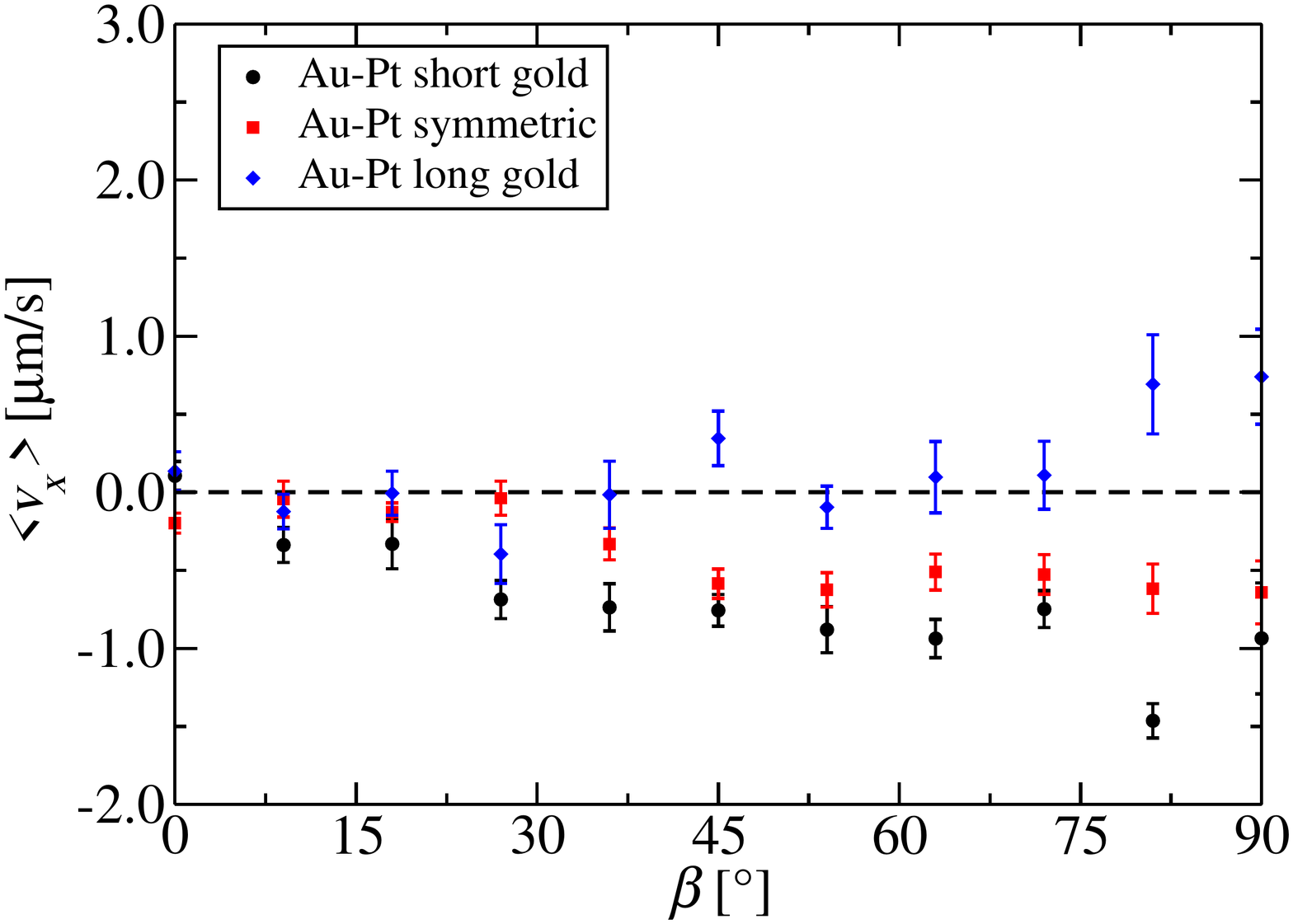}
  \includegraphics[width=0.49 \columnwidth]{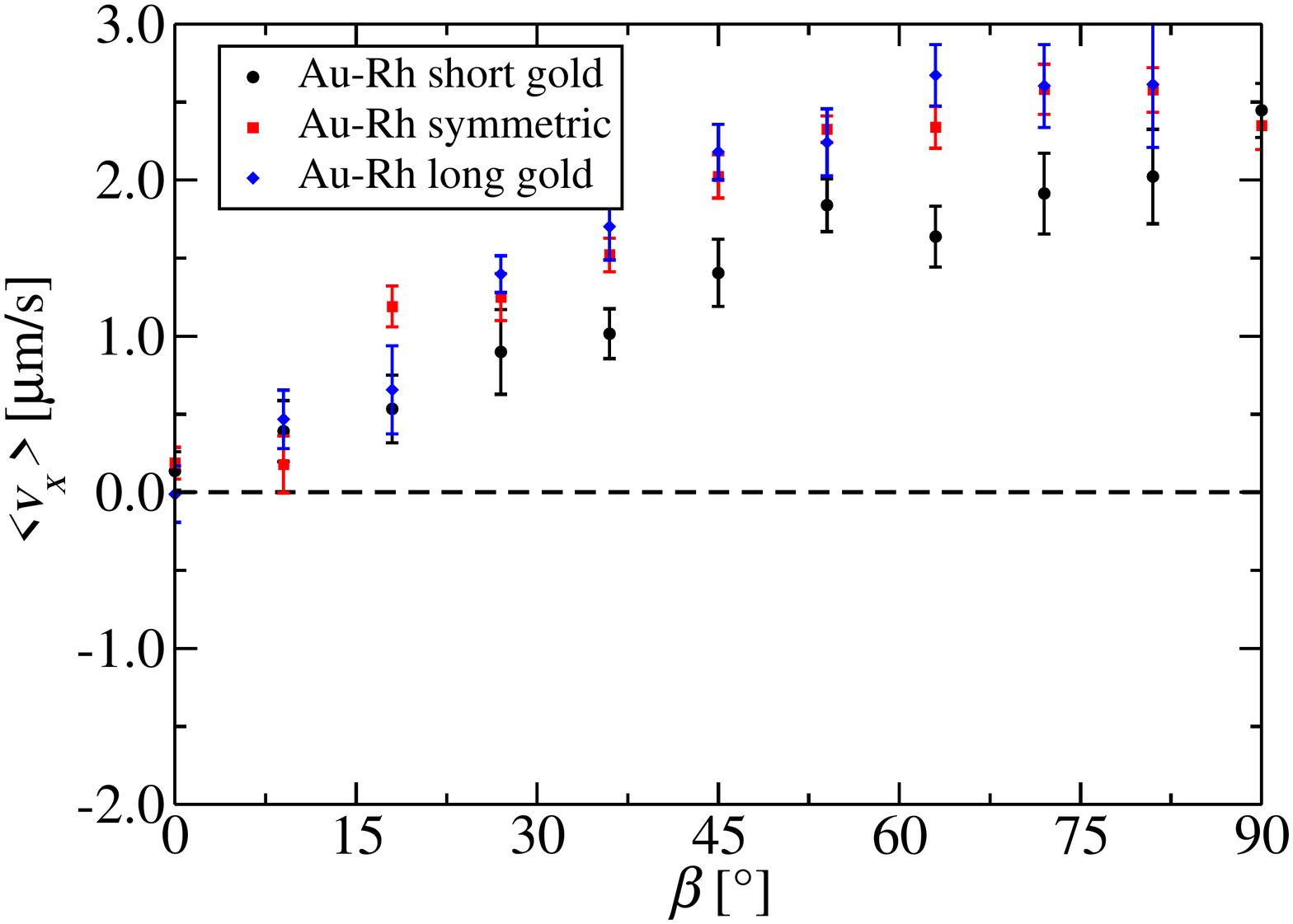}
  \includegraphics[width=0.49 \columnwidth]{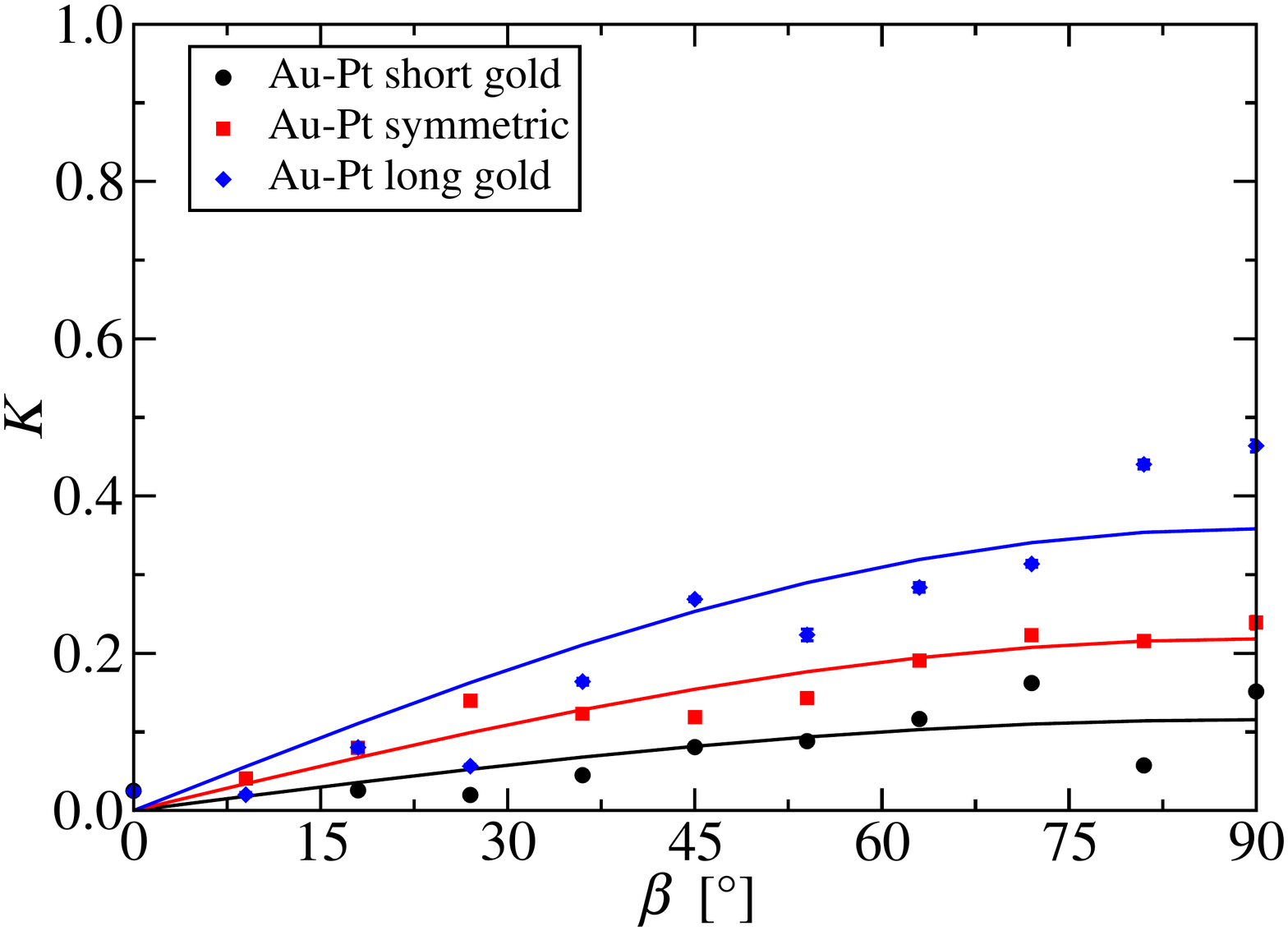}
  \includegraphics[width=0.49 \columnwidth]{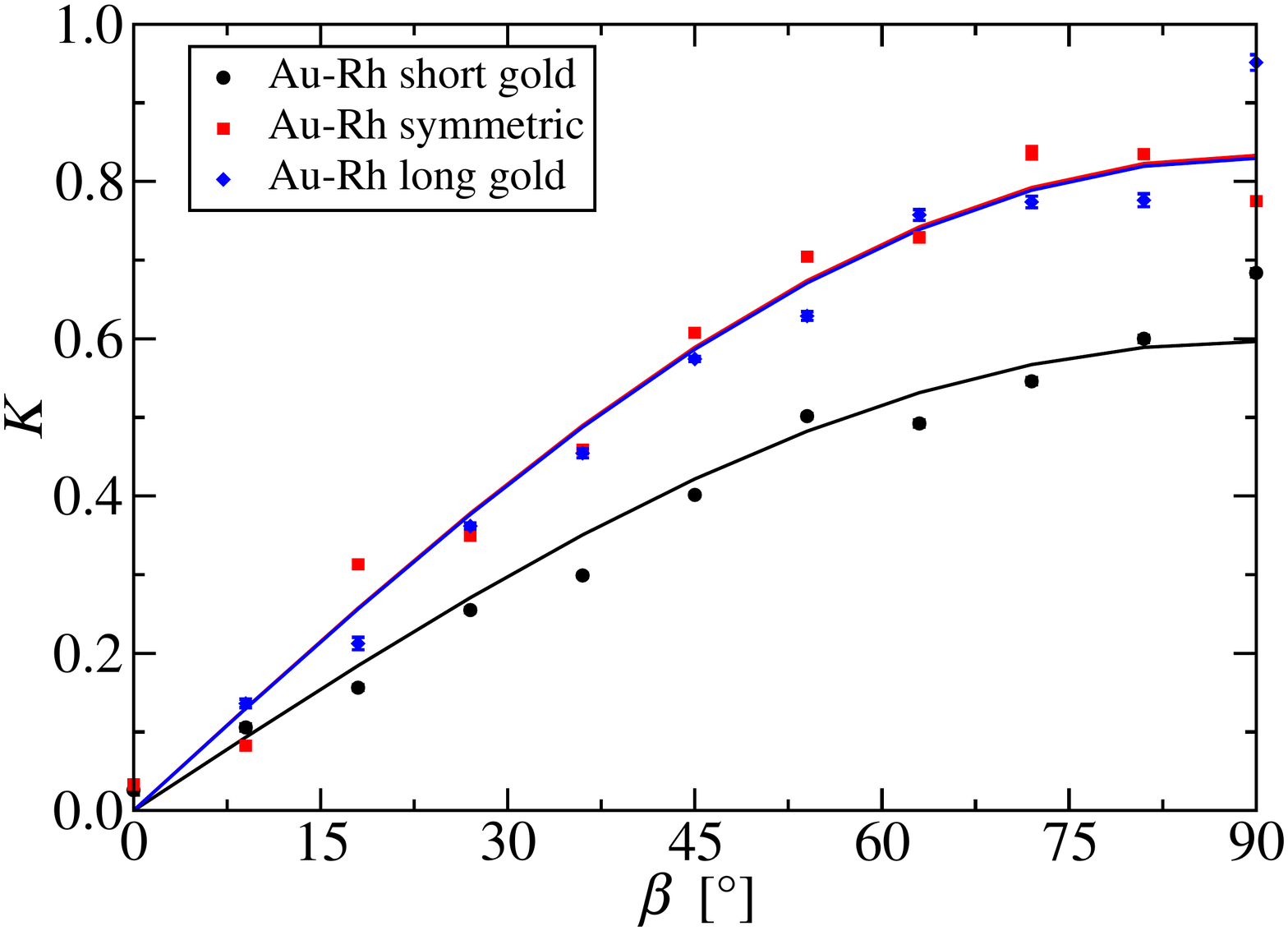}
  \caption{Results from Brownian simulations of $2\, \si{\mu m}$ long rods swimming near a wall.
{\bf Top panels:} mean upward velocity versus wall inclination for Au-Pt (Left) and Au-Rh rods (Right). 
{\bf Bottom panels:} $K$ parameter versus wall inclination for Au-Pt (Left) and Au-Rh rods (Right).}
  \label{fig:vxVsAlpha}
\end{figure}

In the top panels of Fig. \ref{fig:vxVsAlpha}, we show the average velocity along the $x$-axis versus the wall inclination for Au-Pt and Au-Rh rods. In the bottom panels of Fig. \ref{fig:vxVsAlpha}, we show $K$ versus the wall inclination $\beta$.
The Au-Rh rods swim upwards for all inclinations and all swimmer types. Meanwhile, Au-Pt rods with short-gold segement and symmetric swimmers fall although their orientation show that they point upwards most of the time.
Au-Pt rods with long-gold segment show a weak upward swimming bias.
All these results agree qualitatively with the experiments although the numerical swimmers are better gravitactors. 
We note that in the simulations we model the active slip instead of  solving the complicated electrochemical problem that ultimately creates the active flows. 
Moreover, we ignore if there are electrostatic forces between the rods and the wall which could affect the results. 
For these reasons, we do not expect a perfect agreement between the simulations and the experimental results.



\FloatBarrier

\bibliography{Biblio_gravi.bib}